\documentclass[letterpaper]{article}

\usepackage{flairs}
\usepackage{times}
\usepackage{helvet}
\usepackage{courier}
\usepackage[final]{graphicx}
\usepackage{url}
\usepackage{booktabs}
\usepackage{amsmath,amssymb,amsfonts}
\usepackage{algorithmic}
\usepackage{textcomp}
\usepackage{multirow}
\usepackage{array}
\usepackage{ragged2e}
\newcolumntype{L}[1]{>{\RaggedRight\arraybackslash}p{#1}}
\title{InsightBoard: An Interactive Multi-Metric Visualization and Fairness Analysis Plugin for TensorBoard}

\author{Ray Zeyao Chen \and Christan Grant\\
Department of Computer \& Information Science \& Engineering\\
University of Florida, Gainesville, FL, USA\\
\{chenz1, christan\}@ufl.edu
}

\begin{document}
\maketitle

\begin{abstract}
Modern machine learning systems deployed in safety-critical domains require visibility not only into aggregate performance but also into how training dynamics affect subgroup fairness over time. Existing training dashboards primarily support single-metric monitoring and offer limited support for examining relationships between heterogeneous metrics or diagnosing subgroup disparities during training.
We present InsightBoard, an interactive TensorBoard plugin that integrates synchronized multi-metric visualization with slice-based fairness diagnostics in a unified interface. InsightBoard enables practitioners to jointly inspect training dynamics, performance metrics, and subgroup disparities through linked multi-view plots, correlation analysis, and standard group fairness indicators computed over user-defined slices.
Through case studies with YOLOX on the BDD100k dataset, we demonstrate that models achieving strong aggregate performance can still exhibit substantial demographic and environmental disparities that remain hidden under conventional monitoring. By making fairness diagnostics available during training, InsightBoard supports earlier, more informed model inspection without modifying existing training pipelines or introducing additional data stores.
\end{abstract}

\section{Introduction}
TensorBoard is widely used to monitor training, but many debugging questions require correlating heterogeneous signals (e.g., loss, learning rate schedules, gradient distributions) across runs and time. In parallel, responsible deployment requires slice-level auditing: strong aggregate scores can coexist with large disparities across demographic and environmental subgroups \cite{mehrabi2021survey}. Existing fairness toolkits can compute disparities, but they are often disconnected from the training dashboard, often leading to fragmented workflows where fairness checks are performed separately from routine training monitoring\cite{fairlearn2020,aifairness3602018}.

We present \textbf{InsightBoard}, a TensorBoard plugin that integrates multi-run, multi-metric monitoring with interactive slice-based fairness analysis in a single interface. InsightBoard supports synchronized multi-view plots, correlation views, and subgroup fairness indicators, enabling iterative ``slice--inspect--adjust--recheck'' cycles during development. In a YOLOX case study on BDD100k, we show that models with high overall accuracy can still exhibit large subgroup gaps, and that InsightBoard makes these trade-offs explicit early in training.

\textbf{Contributions.} This paper makes two contributions:
(1) System and diagnostic integration: We design and implement InsightBoard, a lightweight TensorBoard plugin that unifies synchronized multi-run, multi-metric visualization with slice-based fairness auditing (including standard disparity indicators and IN/OUT analysis), without requiring additional data stores or modifications to existing training workflows.
(2) Empirical validation: Through case studies on YOLOX with BDD100k and an overhead analysis, we demonstrate the feasibility and practical value of integrating fairness diagnostics directly into the training process.

\section{Related Work}
Object detection and autonomous driving datasets have documented subgroup performance disparities that aggregate metrics can mask \cite{bdd100k,liu2021fairness,zhang2021fairness,buolamwini2018gender}. Fairness objectives and the accuracy--fairness trade-off are well studied in classification \cite{hardt2016equality,chouldechova2017fair}, but real-time monitoring during training is less explored for detection.

Fairness toolkits such as Fairlearn and AI Fairness 360 provide reporting and mitigation workflows \cite{fairlearn2020,aifairness3602018}. TensorFlow Fairness Indicators integrates sliced metrics into TensorBoard-style pipelines, while the What-If Tool supports interactive probing \cite{wexler2019whatif}. These tools support interactive fairness analysis, but do not provide synchronized multi-run training trace comparison or correlated multi-metric views within the standard TensorBoard workflow.

\begin{figure*}[t]
\centering
\includegraphics[width=0.7\textwidth]{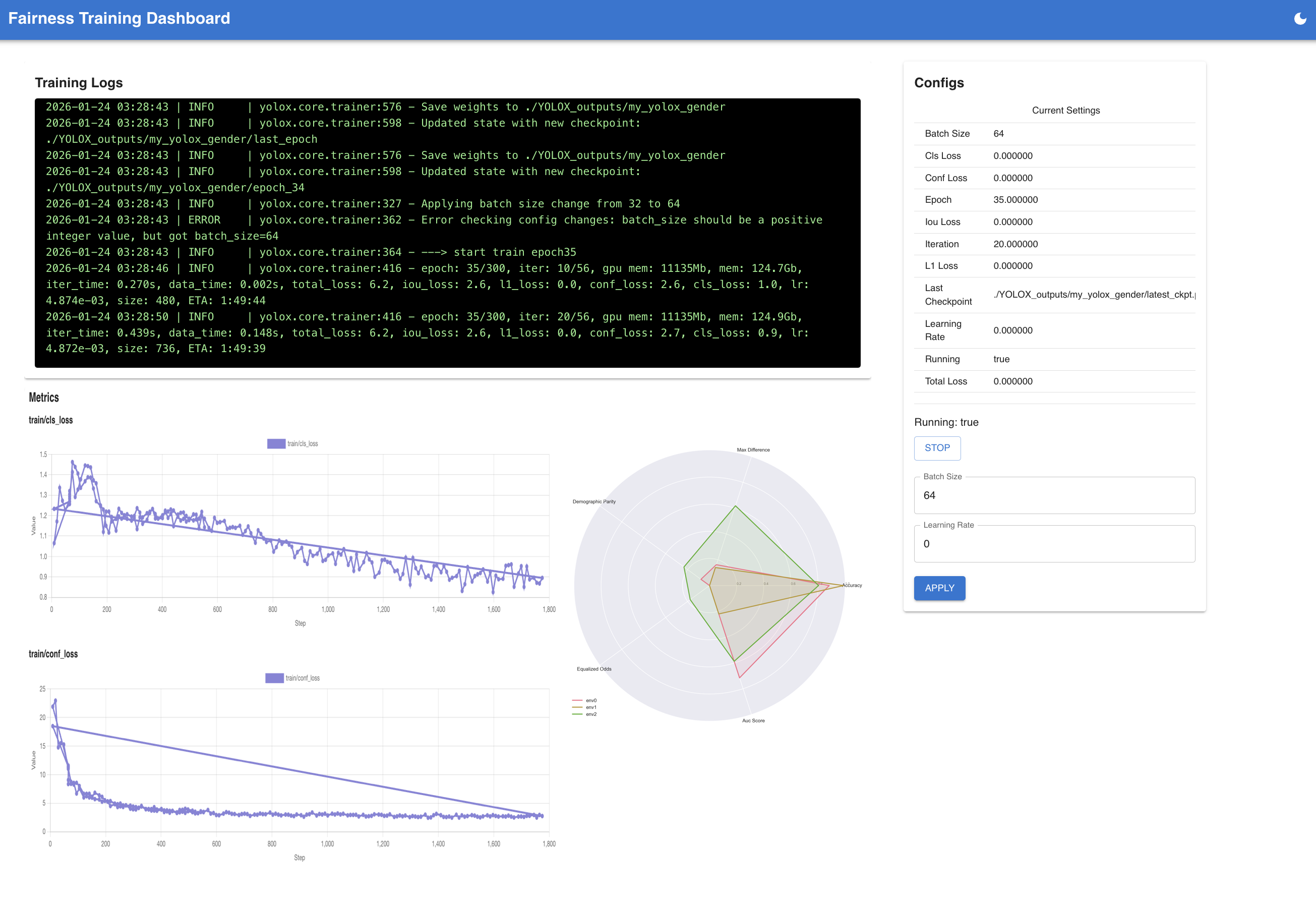}
\alt{System architecture diagram for InsightBoard. The figure shows a TensorBoard-based pipeline with a backend plugin and a frontend dashboard. The backend intercepts TensorBoard data requests, performs on-the-fly metric aggregation and fairness computation from logged training data, and serves processed results to the frontend. The frontend presents coordinated visual views for training logs, metric trends, fairness analysis, and configuration controls. The diagram emphasizes the flow from standard TensorBoard event data through backend processing to interactive visualization.}
\caption{InsightBoard System Architecture. The backend plugin intercepts data requests to perform on-the-fly metric aggregation and fairness calculations, serving this processed data to a new frontend dashboard.}
\label{fig:architecture}
\end{figure*}

Experiment tracking platforms (e.g., Weights \& Biases, MLflow) support multi-run comparisons but do not natively expose fairness indicators or linked correlation views. Conversely, fairness toolkits provide valuable post-hoc auditing but remain disconnected from routine training workflows. InsightBoard bridges this gap by integrating slice-based fairness diagnostics directly into the standard TensorBoard environment. Interactive visual diagnosis has recently proven highly effective in surfacing multidimensional fairness issues \cite{chen2026rise}. However, these approaches are predominantly applied post-deployment. InsightBoard extends the visual diagnostic philosophies of RISE directly into the active training loop, bridging the gap between performance debugging and real-time responsible AI assessment.

\section{System Architecture and Implementation}

InsightBoard is a lightweight, extensible TensorBoard plugin designed to integrate directly into existing MLOps workflows for joint performance and fairness analysis. The key architectural novelty lies in its ability to perform on-the-fly multi-metric synchronization and slice-based fairness computation directly from standard event logs, without requiring secondary databases or custom logging frameworks. To foster community adoption and reproducible research, InsightBoard will be released as an open-source repository on GitHub upon publication. The system follows a modular backend–frontend design built entirely on standard TensorBoard infrastructure (Figure~\ref{fig:architecture}).

\subsection{Backend Processing Pipeline}

The backend operates entirely within the TensorBoard WSGI application state. We utilize the TensorBoard event processing API, specifically the EventMultiplexer, to asynchronously ingest scalar, histogram, and tensor summaries. Because computing intersectional fairness metrics across thousands of steps is memory-intensive, our Metric Aggregator implements a sliding-window reservoir sampling technique. It aligns heterogeneous logs (e.g., learning rate scalars) to a common monotonic axis (training step).

When a user requests a fairness slice, the Fairness Metric Calculator intercepts raw prediction tensors and ground-truth label bindings. It dynamically partitions these arrays based on the requested demographic attributes (e.g., gender, lighting) and computes standard indicators like Demographic Parity Difference and Equalized Odds \cite{hardt2016equality}. This allows the system to serve complex, cross-referenced JSON payloads to the frontend REST endpoints in under 100ms, maintaining the low-latency requirement of interactive dashboards.

\subsection{Frontend Design and Interactive Visualization}

The frontend is a TypeScript and D3.js single-page application packaged as a TensorBoard plugin. It maintains its own client-side state, enabling rich interactions without page reloads. The interface is organized around two tightly integrated views:

\textbf{Multi-Metric Analysis and Correlation.} Users can select multiple experiment runs and arbitrary metric combinations. Visualized on a synchronized x-axis, this enables causal inspection (e.g., relating learning rates to gradient distributions). Interactive brushing highlights corresponding regions across plots, while a correlation heatmap surfaces salient dependencies.

\textbf{Integrated Fairness Analysis.} 
The Fairness Dashboard embeds fairness auditing directly into the training workflow \cite{fairlearn2020,wexler2019whatif}. Users can slice performance by sensitive attributes and inspect disaggregated metrics alongside group fairness indicators (Demographic Parity, Equalized Odds). Furthermore, InsightBoard supports real-time exploration of the accuracy–fairness trade-off through interactive controls. Metrics update live as hyperparameter controls (e.g., fairness weighting terms) change, enabling rapid what-if analysis to identify Pareto-optimal operating points. Group fairness metrics, including Demographic Parity Difference and Equalized Odds Difference, are computed and visualized to make inequities immediately apparent \cite{aifairness3602018,hardt2016equality}.

\begin{figure}[t]
\centering
\includegraphics[width=0.45\textwidth]{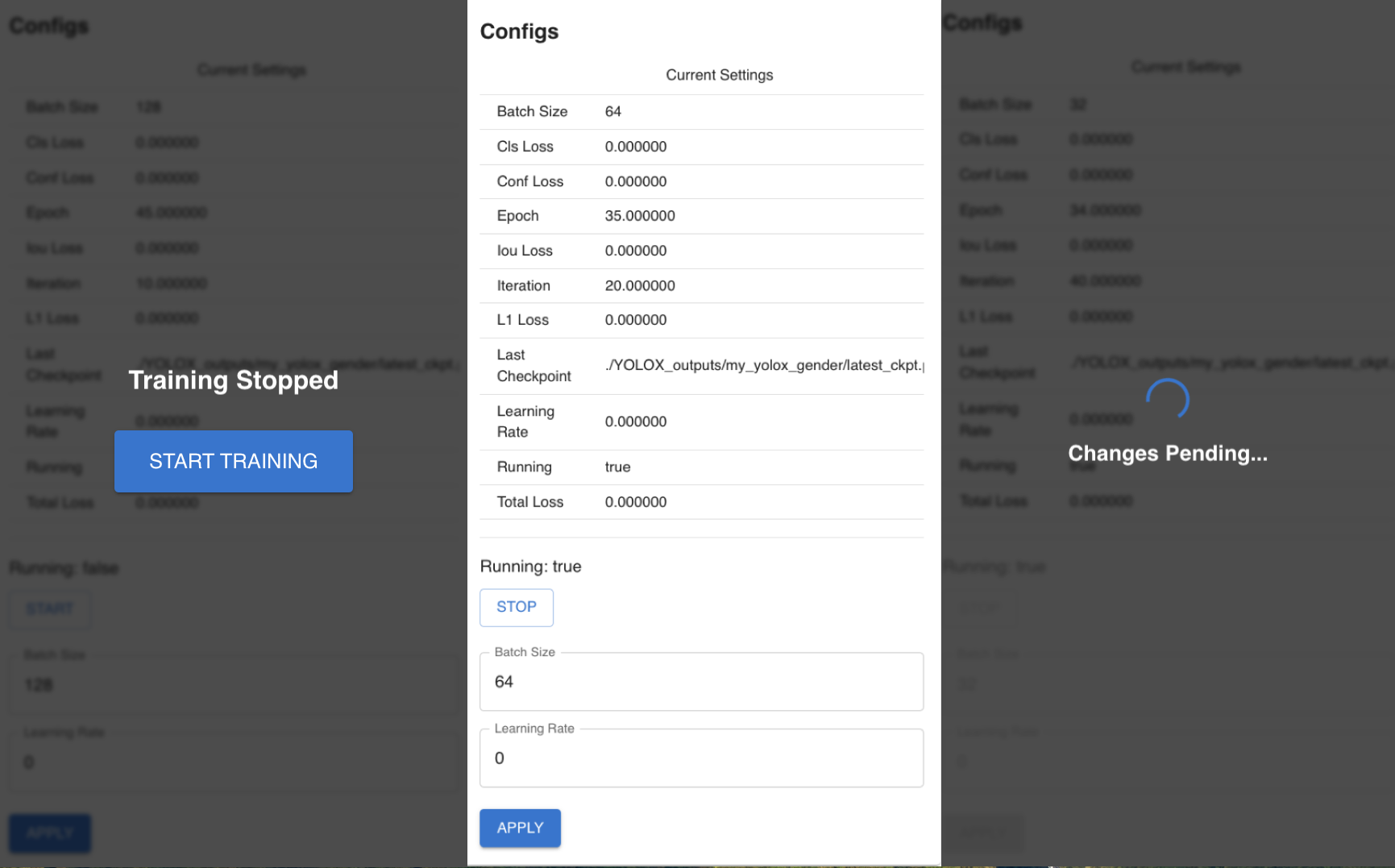}
\alt{Configuration interface for interactive fairness analysis. The figure shows a control panel with editable training and fairness-related hyperparameters on the left and a linked visualization on the right. The interface illustrates how changes to settings such as fairness penalty weights correspond to changes in a fairness metric over training. The display highlights interactive what-if analysis during model development.}
\caption{The configuration interface. The left panel shows specific hyperparameter deltas (e.g., fairness penalty weights), while the linked timeline on the right displays the corresponding real-time impact on the Equalized Odds gap during the training progression.}
\label{fig:config}
\end{figure}

InsightBoard also supports real-time exploration of the accuracy–fairness trade-off through interactive controls (Figure~\ref{fig:config}). 
In our experimental setup, we expose selected training hyperparameters, including a fairness weighting term implemented externally to the model, to support controlled what-if analysis. InsightBoard visualizes the resulting metric changes but does not prescribe or automate optimization strategies.
Metrics update live as controls change, enabling rapid what-if analysis. In BDD100k experiments with YOLOX, this capability revealed persistent gender-based AP gaps even as overall mAP improved, allowing users to identify Pareto-optimal operating points by tuning fairness constraints during training.

% \subsection{Summary and Positioning}
% By unifying multi-metric correlation analysis and in-training fairness monitoring within a single TensorBoard plugin, InsightBoard bridges a gap between performance debugging and responsible AI assessment. As summarized in Table~\ref{tab:comparison}, it extends TensorBoard’s native capabilities while complementing standalone fairness toolkits, offering a cohesive, low-overhead solution for interactive, fairness-aware model development.

\begin{figure}[t]
\centering
\includegraphics[width=0.3\textwidth]{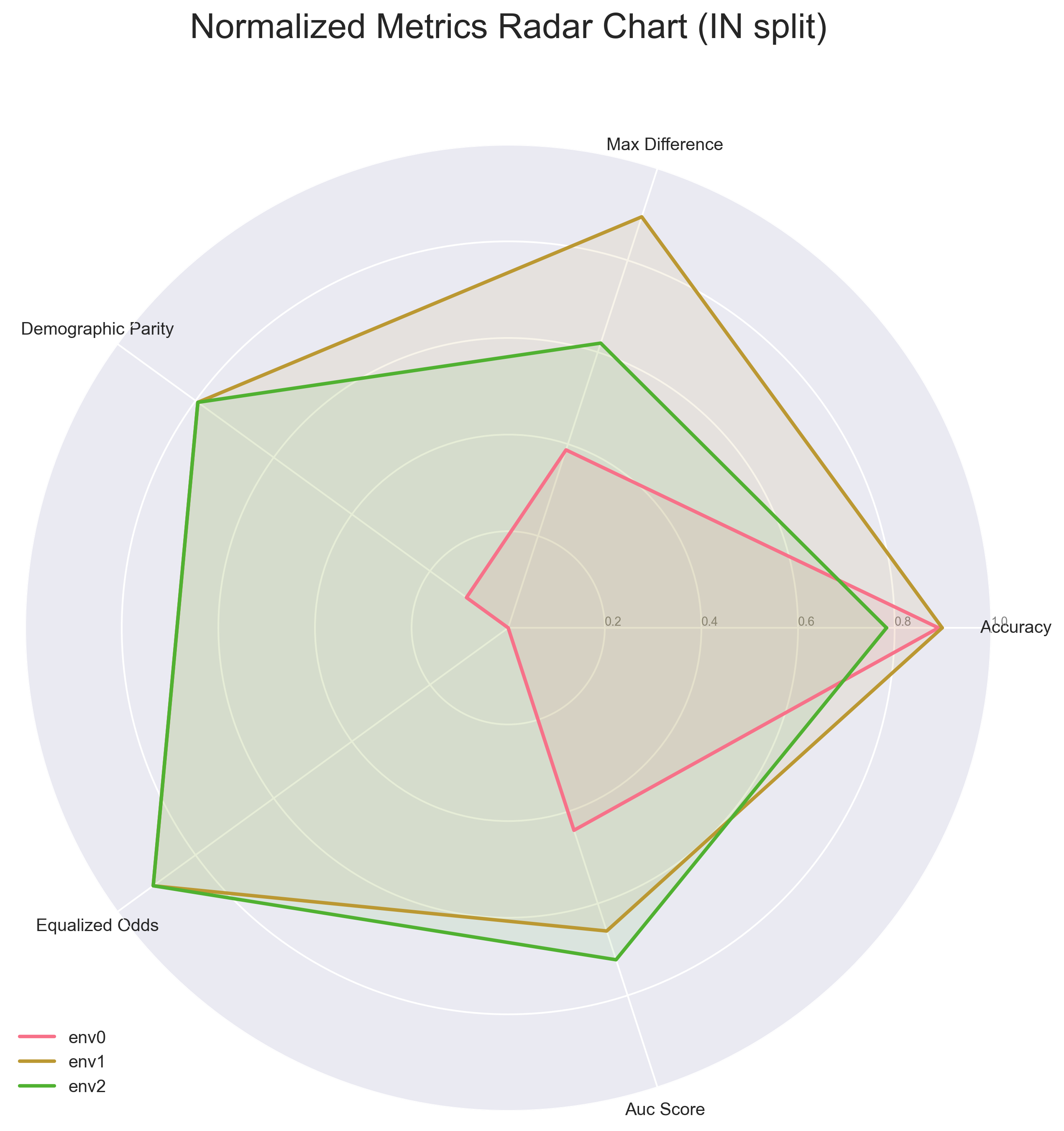}
\alt{Radar chart for IN-distribution fairness evaluation. Multiple polygon overlays compare model runs across subgroup-specific detection performance axes. The axes represent disaggregated mAP for demographic and environmental subgroups, such as male/daytime and female/nighttime. Larger and more balanced polygon shapes indicate stronger and more equitable subgroup performance, while narrow or uneven regions indicate disparity across slices.}
\caption{The Fairness Dashboard visualizing IN-distribution evaluation. The radar chart axes represent disaggregated mAP across specific subgroups (e.g., Male/Daytime vs. Female/Nighttime). The area of the polygon allows practitioners to visually compare the overall equity footprint of competing model runs.}
\label{fig:fairness}
\end{figure}

\begin{table}[t]
\caption{Feature comparison of visualization and fairness tools.}
\label{tab:comparison}
\centering
\small
\begin{tabular}{@{}L{2.1cm}ccc@{}}
\toprule
\textbf{Feature} & \textbf{TensorBoard} & \textbf{Fairlearn} & \textbf{InsightBoard} \\
\midrule
Metric tracking & \checkmark & \textemdash & \checkmark \\
Multi-metric correlation & \textemdash & \textemdash & \checkmark \\
Slice-based fairness & \textemdash & \checkmark & \checkmark \\
In-training monitoring & \textemdash & \textemdash & \checkmark \\
\bottomrule
\end{tabular}
\end{table}

\section{Fairness Metrics and Interaction Design}

\textbf{Disaggregated performance.} InsightBoard supports task-dependent performance metrics. For classification, we track accuracy, precision/recall, and AUC; for detection, we track AP under user-selected IoU thresholds \cite{yolox2021}. All metrics can be computed per-slice (e.g., gender, age, lighting, domain), enabling users to identify where aggregate improvements hide subgroup regressions.

\textbf{Fairness metrics.} We include commonly used group disparity summaries such as Demographic Parity (DP) gap and Equalized Odds (EO) gaps \cite{hardt2016equality}. For binary classification, DP gap can be expressed as:
$\Delta_{\mathrm{DP}} = \left| \Pr(\hat{y}=1 \mid a=0) - \Pr(\hat{y}=1 \mid a=1) \right|.$
EO is captured by disparities in error rates, such as differences in true positive rates and false positive rates across groups. These indicators are intended as diagnostic signals rather than final judgments, consistent with survey guidance on fairness evaluation \cite{mehrabi2021survey}.

\textbf{Distribution stability.} InsightBoard reports metrics across IN (train/val/test) and OUT environments when available, enabling users to distinguish fairness improvements that persist from those that are brittle under distribution shift (Figure~\ref{fig:fairness}).

\textbf{Linked interaction model.} InsightBoard uses cross-view linking, where selecting a run, epoch range, or subgroup in one view highlights the corresponding elements across all views (timelines, subgroup bars, and correlation heatmaps), supporting rapid identification of patterns such as fairness–performance trade-offs. A complementary text pane summarizes configuration deltas (e.g., optimizer, learning rate schedule, batch size) for the selected run.

\section{Experiments and Case Studies}

Our experiments serve as diagnostic case studies to evaluate whether InsightBoard enables practitioners to detect and interpret performance–fairness trade-offs that remain hidden under standard dashboards. We use YOLOX \cite{yolox2021}, a widely adopted real-time object detector, trained on BDD100k, a large-scale autonomous driving dataset containing demographic and environmental annotations (e.g., lighting, weather).

\textbf{Dataset. }
BDD100k is a large-scale autonomous driving dataset with demographic and environmental annotations relevant for fairness analysis \cite{bdd100k}. Its diversity across lighting and weather conditions enables slice-level evaluation of detection performance.
\textbf{Model. }
We use YOLOX \cite{yolox2021}, a widely adopted real-time object detector, as a representative modern detection model.”

% \subsection{Accuracy-Fairness Trade-off}
% A fundamental challenge in fair machine learning is the tension between accuracy and fairness objectives. When optimizing solely for accuracy, models may learn and amplify biases present in the training data, leading to discriminatory outcomes. For instance, if a dataset contains more daytime images with male pedestrians, a model optimized for accuracy may achieve higher detection rates for this subgroup while underperforming on underrepresented groups (e.g., female pedestrians in nighttime conditions). Conversely, optimizing for fairness may require sacrificing some aggregate accuracy to ensure more equitable performance across groups. InsightBoard supports exploration of accuracy–fairness trade-offs by visualizing how metric changes coincide with training dynamics.

% \subsection{Pilot Case Study: Multi-Attribute Classification}
% In addition to the BDD100k pedestrian detection experiments, we conducted a pilot study on a multi-attribute classification task. This dataset configuration predicts a three-class skin tone label ($skin_0 / skin_1 / skin_2$), with gender as a sensitive attribute and age as an environment attribute. The distribution is imbalanced (e.g., $skin_0$: 6,614; $skin_1$: 2,883; $skin_2$: 847; adult: 6,293; young: 3,051), motivating slice-level reporting and careful selection of disparity summaries. Table~\ref{tab:dataset} summarizes the dataset composition.

\begin{table}[t]
\centering
\small
\caption{Pilot dataset composition reported during development.}
\label{tab:dataset}
\begin{tabular}{@{}L{2.0cm}L{2.2cm}r@{}}
\toprule
\textbf{Attribute} & \textbf{Category} & \textbf{Count} \\
\midrule
Skin tone & $skin_0$ (unknown) & 6614 \\
 & $skin_1$ (light) & 2883 \\
 & $skin_2$ (dark) & 847 \\
\midrule
Age & $age_0$ (adult) & 6293 \\
 & $age_1$ (young) & 3051 \\
\bottomrule
\end{tabular}
\end{table}

\subsection{Implementation Details}
InsightBoard is implemented as a Python backend plugin and a TypeScript frontend application. The backend leverages TensorBoard's plugin architecture to register custom routes that intercept data requests, perform on-the-fly metric computation, and serve processed data via REST APIs. The frontend uses D3.js for visualization rendering and React for state management, enabling complex interactions such as linked brushing and real-time updates without page reloads.

For YOLOX integration, we extended the training loop to log fairness metrics alongside standard performance metrics. Specifically, we compute subgroup-specific metrics by filtering predictions and ground truth annotations based on demographic attributes from BDD100k. These metrics are logged as TensorBoard scalar summaries, which InsightBoard then aggregates and visualizes. The fairness weight parameter is implemented as a hyperparameter that scales the contribution of fairness loss terms in the training objective, allowing real-time exploration of the accuracy-fairness trade-off.

\subsection{System Overhead Evaluation}
We evaluated InsightBoard's computational overhead across different scales. For a typical YOLOX training run with 100 epochs, parsing event logs takes approximately 2-3 seconds. Metric computation time per epoch is negligible ($<$10ms) for standard metrics, but increases to 50-100ms when computing fairness metrics across multiple slices (e.g., gender $\times$ lighting condition). Frontend responsiveness remains acceptable (refresh latency $<$500ms) for up to 50 concurrent runs, degrading gracefully beyond this scale. These overheads are minimal compared to training time, making real-time fairness monitoring feasible without significant performance impact.

% \subsection{User Study Design}
% To further evaluate InsightBoard’s impact on fairness debugging, we outline a controlled user study design comparing standard TensorBoard workflows with TensorBoard augmented by InsightBoard. The proposed tasks focus on identifying worst-performing slices, interpreting performance–fairness trade-offs, and tracing configuration changes that induce fairness regressions. While this study is left to future work, it provides a structured framework for evaluating the effectiveness of integrated fairness diagnostics in realistic development settings.

\begin{table}[t]
\centering
\small
\caption{Evaluation tasks and outcome measures for a controlled comparison study.}
\begin{tabular}{@{}L{2.6cm}L{4.4cm}@{}}
\toprule
\textbf{Task} & \textbf{Outcome measure} \\
\midrule
Find worst slice & correctness; time \\
Explain trade-off & explanation quality; time \\
Pick best run & correctness; confidence \\
Locate config cause & correctness; time \\
\bottomrule
\end{tabular}
\label{tab:tasks}
\end{table}

We demonstrate InsightBoard's capabilities through two core scenarios using YOLOX models trained on the BDD100k dataset. These scenarios illustrate how InsightBoard enables practitioners to identify and analyze fairness issues that are invisible when examining only aggregate performance metrics.

\subsection{Case Study: Correlation-Driven Debugging and Fairness Diagnosis}

We demonstrate InsightBoard on YOLOX trained on BDD100k, focusing on how integrated multi-metric and slice-based analysis reveals both training instabilities and hidden subgroup disparities.

First, comparing two runs with different learning rates (0.01 vs. 0.001), InsightBoard’s synchronized multi-metric views (mAP, loss, learning rate, and gradient norm distributions) reveal that the higher learning rate induces unstable gradient behavior during critical training phases. This instability explains why the model converges faster but achieves lower final mAP (78\% vs. 85\%), providing a direct link between hyperparameter choice, internal dynamics, and performance.

Second, slice-based analysis exposes substantial subgroup disparities masked by aggregate metrics. While a baseline model achieves 85.2\% overall mAP, InsightBoard reveals a large gender–environment gap (94.8\% for male/daytime vs. 59.3\% for female/nighttime). Notably, this divergence becomes visible early in training (around epoch 20). By adjusting the fairness penalty and applying targeted augmentation, we observe a trade-off in real time: aggregate mAP decreases slightly (to 83.1\%), while the subgroup gap is substantially reduced (from 35\% to 18\%).

Together, these results show that InsightBoard enables early detection of both optimization issues and fairness failures, supporting more informed and efficient model iteration.

\subsection{Key Findings}

Our results demonstrate that aggregate metrics consistently mask intersectional failures in safety-critical object detection. InsightBoard advances the subject domain by collapsing the "train, then audit" lifecycle into a single, cohesive workflow. By exposing these gaps dynamically, practitioners can halt biased models early, saving compute resources, and actively steer the model toward a more equitable Pareto frontier during hyperparameter tuning rather than attempting post-hoc mitigation.

\textbf{Precision vs. Recall Trade-off:} In safety-critical applications like autonomous driving, false negatives (missed pedestrians) are typically more costly than false positives (false alarms). Therefore, practitioners may prioritize recall over precision, accepting more false alarms to ensure pedestrian safety. InsightBoard enables real-time exploration of this trade-off by visualizing precision and recall disaggregated by demographic groups, revealing whether precision-recall imbalances disproportionately affect certain subgroups.

\textbf{Accuracy-Fairness Trade-off:} Perhaps the most fundamental trade-off is between aggregate accuracy and fairness. Optimizing solely for accuracy (e.g., maximizing $cocoAP5$) may lead models to exploit demographic correlations present in training data, resulting in biased outcomes. Conversely, enforcing strict fairness constraints may reduce aggregate performance. InsightBoard's real-time parameter adjustment allows practitioners to explore the accuracy-fairness Pareto frontier, identifying configurations that achieve acceptable balances for their specific deployment context.

\textbf{Effectiveness of Visual Representations}
Rather than introducing novel chart types, InsightBoard's value stems from the tight spatial integration of standard visual tools. The correlation matrix serves as a standard utility to quickly surface inverse relationships between aggregate accuracy and fairness gaps. By linking these standard views (timelines, radar charts, and matrices) to a shared state, users can immediately correlate a spike in learning rate to a degradation in a specific demographic slice's performance. Furthermore, radar charts (Figure~\ref{fig:fairness}) enable users to assess whether fairness improvements persist across IN/OUT data distributions—a critical consideration for real-world deployment.

\section{Conclusion}

We presented InsightBoard, an interactive TensorBoard plugin that integrates multi-metric correlation analysis with slice-based fairness diagnostics during training. Through case studies on YOLOX and BDD100k, we showed that strong aggregate performance can coexist with substantial demographic and environmental disparities that remain invisible under conventional monitoring.

By embedding fairness diagnostics directly into the training workflow, InsightBoard enables earlier and more informed inspection of performance–fairness trade-offs. This capability is particularly valuable in safety-sensitive domains where late discovery of subgroup failures carries serious consequences. Future work includes extending InsightBoard to additional model classes (e.g., language models), such as MultiScript30k \cite{driggers2025multiscript30k}, evaluating its applicability on multi-script datasets to study non-vision fairness diagnostics, and conducting comprehensive user studies to empirically validate its interaction design.

\bibliographystyle{flairs}  
\bibliography{bibliography}

@article{yolox2021,
  title        = {YOLOX: Exceeding YOLO Series in 2021},
  author       = {Ge, Zheng and Liu, Songtao and Wang, Feng and Li, Zeming and Sun, Jian},
  journal      = {arXiv preprint arXiv:2107.08430},
  year         = {2021}
}

@article{mehrabi2021survey,
  title        = {A Survey on Bias and Fairness in Machine Learning},
  author       = {Mehrabi, Ninareh and Morstatter, Fred and Saxena, Nripsuta and Lerman, Kristina and Galstyan, Aram},
  journal      = {ACM Computing Surveys},
  volume       = {54},
  number       = {6},
  pages        = {1--35},
  year         = {2021}
}

@techreport{fairlearn2020,
  title        = {Fairlearn: A Toolkit for Assessing and Improving Fairness in {AI}},
  author       = {Bird, Sarah and Dudik, Miroslav and Edgar, Richard and Horn, Brandon and Lutz, Roman and Milan, Vanessa and Sameki, Mehrnoosh and Wallach, Hanna and Walker, Kathleen and others},
  institution  = {Microsoft},
  number       = {MSR-TR-2020-32},
  year         = {2020}
}

@article{aifairness3602018,
  title        = {AI Fairness 360: An Extensible Toolkit for Detecting, Understanding, and Mitigating Unwanted Algorithmic Bias},
  author       = {Bellamy, Rachel K. E. and Dey, Kuntal and Hind, Michael and Hoffman, Samuel C. and Houde, Stephanie and Kannan, Kalapriya and Lohia, Pranay and Martino, Julio and Mehta, Sameep and Mojsilovic, Aleksandra and others},
  journal      = {arXiv preprint arXiv:1810.01943},
  year         = {2018}
}

@inproceedings{hardt2016equality,
  title        = {Equality of Opportunity in Supervised Learning},
  author       = {Hardt, Moritz and Price, Eric and Srebro, Nati},
  booktitle    = {Advances in Neural Information Processing Systems},
  year         = {2016}
}

@article{wexler2019whatif,
  title        = {The What-If Tool: Interactive Probing of Machine Learning Models},
  author       = {Wexler, James and Pushkarna, Mahima and Bolukbasi, Tolga and Wattenberg, Martin and Viegas, Fernanda and Wilson, Jimbo},
  journal      = {IEEE Transactions on Visualization and Computer Graphics},
  volume       = {26},
  number       = {1},
  pages        = {56--65},
  year         = {2019}
}

@inproceedings{bdd100k,
  title        = {BDD100k: A Diverse Driving Dataset for Heterogeneous Multitask Learning},
  author       = {Yu, Fisher and Xian, Wenqi and Chen, Yingying and Liu, Fangchen and Liao, Mike and Madhavan, Vashisht and Darrell, Trevor},
  booktitle    = {Proceedings of the IEEE/CVF Conference on Computer Vision and Pattern Recognition},
  pages        = {2636--2645},
  year         = {2020}
}

@inproceedings{buolamwini2018gender,
  title        = {Gender Shades: Intersectional Accuracy Disparities in Commercial Gender Classification},
  author       = {Buolamwini, Joy and Gebru, Timnit},
  booktitle    = {Proceedings of the Conference on Fairness, Accountability, and Transparency},
  pages        = {77--91},
  year         = {2018}
}

@article{liu2021fairness,
  title        = {Fairness in Deep Learning: A Computational Perspective},
  author       = {Liu, Lydia T and Dean, Sarah and Rolf, Esther and Simchowitz, Max and Hardt, Moritz},
  journal      = {IEEE Security \& Privacy},
  volume       = {19},
  number       = {5},
  pages        = {81--89},
  year         = {2021}
}

@inproceedings{zhang2021fairness,
  title        = {Fairness in Object Detection: A Survey},
  author       = {Zhang, Han and Lu, Yifei and Wang, Yuchen and others},
  booktitle    = {Proceedings of the IEEE/CVF Conference on Computer Vision and Pattern Recognition Workshops},
  year         = {2021}
}

@article{chouldechova2017fair,
  title        = {Fair Prediction with Disparate Impact: A Study of Bias in Recidivism Prediction Instruments},
  author       = {Chouldechova, Alexandra},
  journal      = {Big Data},
  volume       = {5},
  number       = {2},
  pages        = {153--163},
  year         = {2017}
}

@article{driggers2025multiscript30k, title={MultiScript30k: Leveraging Multilingual Embeddings to Extend Cross Script Parallel Data}, author={Driggers-Ellis, Christopher and Brinkley, Detravious and Chen, Ray and Dhawan, Aashish and Wang, Daisy Zhe and Grant, Christan}, journal={arXiv preprint arXiv:2512.11074}, year={2025} }

@article{chen2026rise,
  title={RISE: Interactive Visual Diagnosis of Fairness in Machine Learning Models},
  author={Chen, Ray and Grant, Christan},
  journal={arXiv preprint arXiv:2602.04339},
  year={2026}
}

\end{document}